\documentclass[twocolumn,prl,aps,floatfix,epsf,psfig,groupaddress,amssymb]{revtex4}
\usepackage{graphicx,subfigure}
\usepackage{epsfig}
\usepackage{mathtools}
\DeclarePairedDelimiter{\evdel}{\langle}{\rangle}
\newcommand{\ev}{\operatorname{}\evdel}
\usepackage{amsmath}
\usepackage{float}
\begin{document}
\title{Magnetic Phases of Graphene Nanoribbons under Potential Fluctuations}
\author{H. U. \"{O}zdemir}
\author{A. Alt{\i}nta\c{s}}
\author{A. D. G\"u\c{c}l\"u}

\affiliation{Department of Physics, Izmir Institute of Technology, IZTECH,
  TR35430, Izmir, Turkey}
%\affiliation{Department of Physics, Izmir Institute of Technology, IZTECH,
%  TR35430, Izmir, Turkey}

\date{\today}

\begin{abstract}
We investigate the effects of long-range potential fluctuations and
electron-electron interactions on electronic and magnetic properties
of graphene nanoribbons with zigzag edges using an extended mean-field
Hubbard model. We show that electron-electron interactions make the
edge states robust against potential fluctuations. When the disorder
is strong enough, the presence of electron-hole puddles induces a
magnetic phase transition from antiferromagnetically coupled edge
states to ferromagnetic coupling, in agreement with recent
experimental results.
\end{abstract}
\maketitle

\section{Introduction}
Graphene\cite{Neto2009,Novoselov2004}, a two-dimensional honeycomb
lattice of carbon atoms, has been the subject of intense investigation
for nanoelectronic and spintronic applications due to its high
electric and thermal conductivity, and intrinsic
magnetism\cite{Son2006a,Wimmer2008,Bundesmann2013,Guclu+09,Fernandez-Rossier2007,GucluWigner}. 
Although pure graphene is not
expected to be magnetic, if the sublattice symmetry of the honeycomb
lattice is broken, there is a possibility to induce
magnetism\cite{Lieb1989}. In particular, atomic-scale engineered
graphene nanoribbons with zigzag orientation are expected to exhibit
magnetized edges with antiferromagnetic coupling between the opposite
edges as confirmed by a large number of theoretical
literature\cite{Fujita,Nakada1996,Wakabayashi+98,Yazyev+08,Wunsch+08,Yamashiro2003,Yazyev+PRB+11,Feldner2010,Wang2009,Cao+2013,Jaskolski+15,Carvalho} in agreement with Lieb's theorem\cite{Lieb1989}. However,
most likely due to limited control over edge structure in real
applications, direct experimental observation is still lacking.
Recently, a semiconductor to metal transition as a function of ribbon
width was observed in nanotailored graphene ribbons with zigzag
edges\cite{Magda2014}. This transition is attributed to a magnetic
phase transition from the antiferromagnetic configuration to the
ferromagnetic configuration, raising hopes for the fabrication of room
temperature graphene-based spintronic devices.

The observation of a magnetic phase transition in graphene nanoribbon
is a surprising result due to the experimental difficulties for
fabricating clean nanostructures with properly passivated and
well-defined
edges\cite{Zhang2013a,Kunstmann2011,Koskinen2008,Koskinen2009,Wassmann2008},
and free from imperfections in the lattice or in the substrate. A
possible source of irregularity in a graphene structure is the
formation of the so called electron-hole
puddles\cite{Bundesmann2013,Rossi2008,DasSarma2011,Schubert+12}. Those
highly inhomogeneous charge distributions were observed by Martin
\textit{et al.}\cite{Martin2007} by mapping the charge neutrality
point. Later Crommie \textit{et al.}\cite{Zhang2009} reported that
impurities between substrate and graphene sheet induce distorted
electron liquid which is in agreement with earlier theoretical works
as well\cite{Rossi2008,Polini2008}. A different study stated that
corrugations are the mechanism behind the formation of charge
inhomogeneities\cite{Gibertini2012}. On the other hand, it was
predicted from tight-binding calculations that the presence of
electron-puddles can mask Anderson localization effects favoring
metallic behavior\cite{Schubert+12}.

In this work, we investigate the effect of electron-hole puddles
resulting from a long-range potential fluctuation on the edge
magnetism of finite nanoribbons, using extended mean-field Hubbard
calculations. We show that, electron-electron interactions increase
the robustness of edge states against disorder as compared to
tight-binding approach in finite graphene nanoribbons. More
importantly, a transition from antiferromagnetic to ferromagnetic edge
phase is observed as the strength of the disorder is increased. These
results are consistent with recent experimental observation of
semiconductor to metal transition as a function of nanoribbon
width\cite{Magda2014}.

This paper has the following structure. In Section II, we introduce
the Hamiltonian model describing the nanoribbon system under
investigation, electron-electron interactions, and the long-range
potential fluctuation. In Section III, the results including the
effects of disorder potential on the electronic properties within the
tight-binding and mean-field models are discussed. The
antiferromagnetic-ferromagnetic phase transition is investigated in
detail for different disorder configurations and interaction
strength. Section IV contains a brief summary.
\begin{figure}
\includegraphics[width=3.3in]{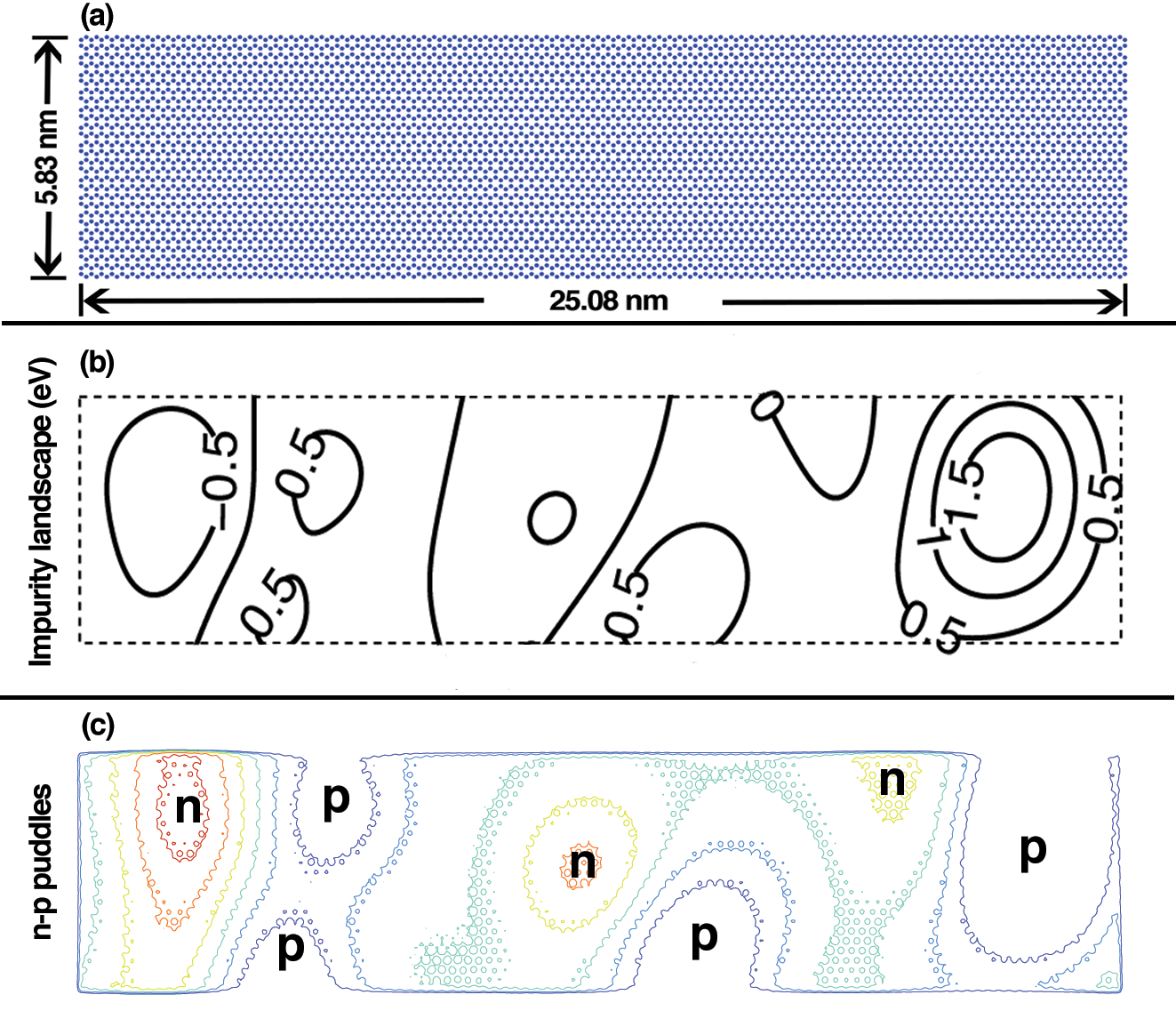}
%\hspace*{-1cm}
%\includegraphics[scale=0.16]{IMP_CONTOUR}
\caption{(Color online) (a) Graphene nanoribbon lattice structure. (b)
  Randomly generated impurity potential landscape.  (c) Total electron
  density showing the formation of electron-hole puddles (regions
  denoted by n and p), obtained from mean-field Hubbard calculations}
\end{figure}

\section{Method and Model}
Our starting point is a single-band tight-binding model for $p_z$ orbitals,
where $s$, $p_x$ and $p_y$ orbitals are neglected as they mainly
contribute to mechanical properties of graphene. Within the mean-field
extended Hubbard model, the Hamiltonian is constructed as follows:
\begin{align}
H_{MFH} =& \sum_{{i,j},\sigma} ( t_{ij} c^{\dagger}_{i,\sigma} c_{j,\sigma} + h.c) \nonumber \\
&+ U\sum_{i} (\ev{n_{i,\uparrow}} - \frac{1}{2})n_{i\downarrow} + (\ev{n_{i,\downarrow}} - \frac{1}{2})n_{i\uparrow} \nonumber\\
&+ \sum_{i,j} V_{ij} (\ev{n_{j}-1}n_{i\downarrow} + \ev{n_j-1}n_{i\uparrow}) \nonumber \\ 
&+\sum_{i\sigma}V_{imp}(i) c^{\dagger}_{i\sigma}c_{i\sigma} 
\end{align}

First term corresponds to tight-binding approximation where the
hopping parameters $t_{ij}$ are taken to be $t_{nn}=-2.8$ eV for
nearest neighbours and $t_{nnn}=-0.1$ eV for next
nearest-neighbours\cite{Neto2009}. The operators
$c^{\dagger}_{i,\sigma}$ and $c_{i,\sigma}$ create and annihilate an
electron at the $i$th orbital with spin $\sigma$,
respectively. The terms $\ev{n_{i,\sigma}}$ denote the
expectation value of electron densities. The second and third terms
are onsite and long range Coulomb interaction terms
respectively. Here, $U$ is taken to be $16.522/ \kappa$ eV, where
$\kappa$ is an effective dielectric constant taken to be as a control
parameter. The long-range interaction parameters $V_{ij}$ are taken to
be $8.64/ \kappa$ eV and $5.33/ \kappa$ eV (Coulomb matrix elements
are calculated numerically by using Slater $\pi_z$ orbitals
\cite{Vij_NUM}) for the first two neighbors, and $1/d_{ij} \kappa$ for
distant neighbors.  $V_{imp}(i)$ represents a smooth long-ranged
potential fluctuation which can be attributed to charge impurities in
the substrate.

Our finite structure contains 5740 atoms respectively giving rise to
about 60 edge states. Length of the lattice vectors are $|\vec{a}_{1,2}|
= 0.151$ nm. The total length of the ribbon is 25.08 nm and the width
is 5.83 nm as shown in Fig. 1a.  The mean-field Hamiltonian is solved
self-consistently in the subspaces of z-component of the total spin
$S_z=(n_{\uparrow}-n_{\downarrow})/2$ (by fixing the number of up and down electrons), and the calculations are
performed for several different $S_z$ values (see for instance Fig.5)
in order to find the ground state magnetic configuration. Each
calculation was repeated several times starting from different initial
density matrices to ensure the convergence to a global energy minimum.

Modelling of the long-range electron-hole puddle disorder that can be
attributed to charged impurities on structure are carried out with a
superposition of Gaussian electrostatic potentials $V_{imp}$ which are
randomly distributed over the sample, creating a smooth potential
landscape (see Fig. 1b). Impurity potential is given by
\begin{equation} 
%\scalebox{1.5}{$V_{imp} = \sum_i V_i \ e^{-\frac{(\vec{r}-\vec{r}_i)^2}{2\sigma^2}}$}
V_{imp}(i) = \sum_n V_n \ e^{-\frac{(\vec{r}_i-\vec{r}_n)^2}{2\sigma^2}}
\end{equation} 
$V_n$ is the potential peak value (randomly chosen between a minimum
and a maximum value) of the $n$th impurity located at $\vec{r}_n$,
$\sigma$ is the width of the potential which is taken to be 10 times
the lattice constant for this study\cite{Zhang2009}. For such
long-ranged scatterers, Anderson localization effects are expected to
be suppressed due to the formation of electron-hole
puddles\cite{Schubert+12}. For all calculations a total of 16 impurity
sources are used, and the calculations are repeated for randomly
generated configurations. Figure 1c shows the formation of
electron-hole puddles (i.e. negatively and positively charged regions)
in the system calculated by subtracting the positive background charge
from the total mean-field electron density.

%whose value effects the type magnetic
%ordering along the edges. Interedge magnetic tail interactions on
%ZGNRs causes ground state to be antiferromagnetic(AFM). However, if
%the width of the ribbons is greater than 7 nm, ground state shows
%ferromagnetic(FM) property due to lack of magnetic tail
%interaction\cite{Magda2014}. We considered these two magnetic states
%by changing difference between spin up and down electron numbers up to
%edge state population. Antiferromagnetic calculations are conducted
%with equal number of spin up and spin down electrons. However,
%ferromagnetic state is modeled with $n_{up}-n_{down}=60$ where all of
%the edge electron aligned in the same direction.

\section{Results}

\begin{figure}
\includegraphics[width=\linewidth]{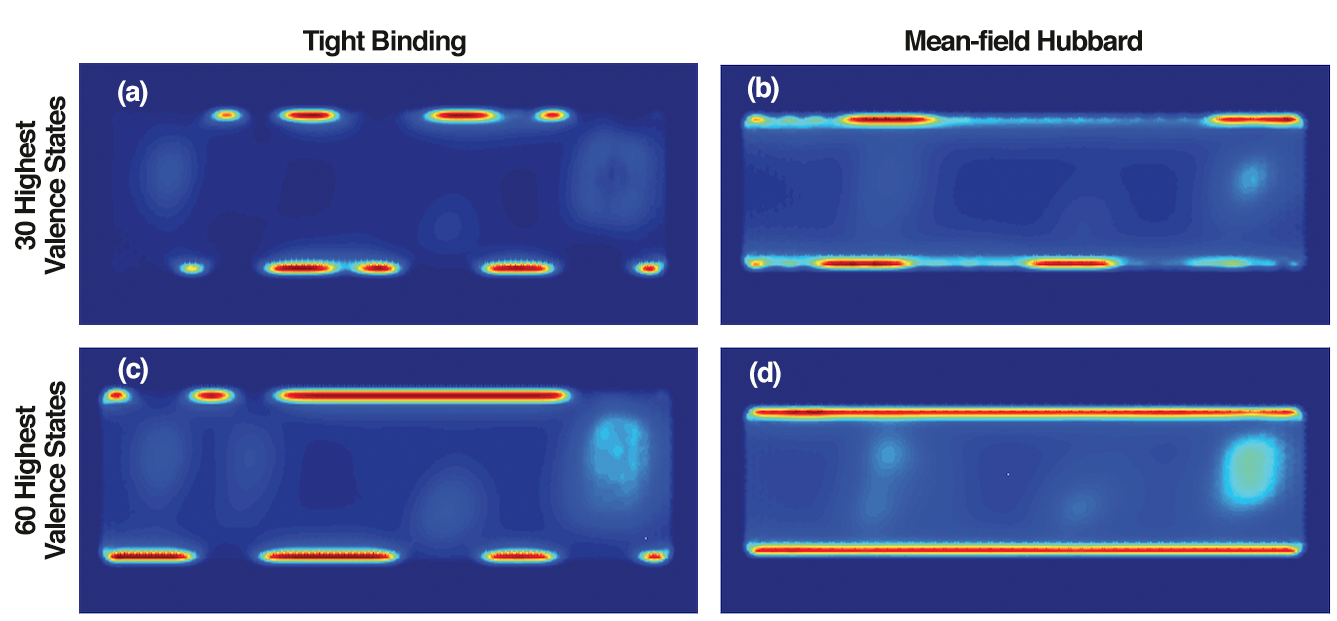}
\caption{(Color online) Electronic density profile corresponding to
the 30 highest occupied valence states (top panels), and the 60 highest occupied valence states (bottom panels), obtained using
tight-binding (left panels) and mean-field Hubbard calculations (right
panels). Electron-electron interactions restore the edge states.}
\end{figure}

\begin{figure*}
\includegraphics[width=\textwidth]{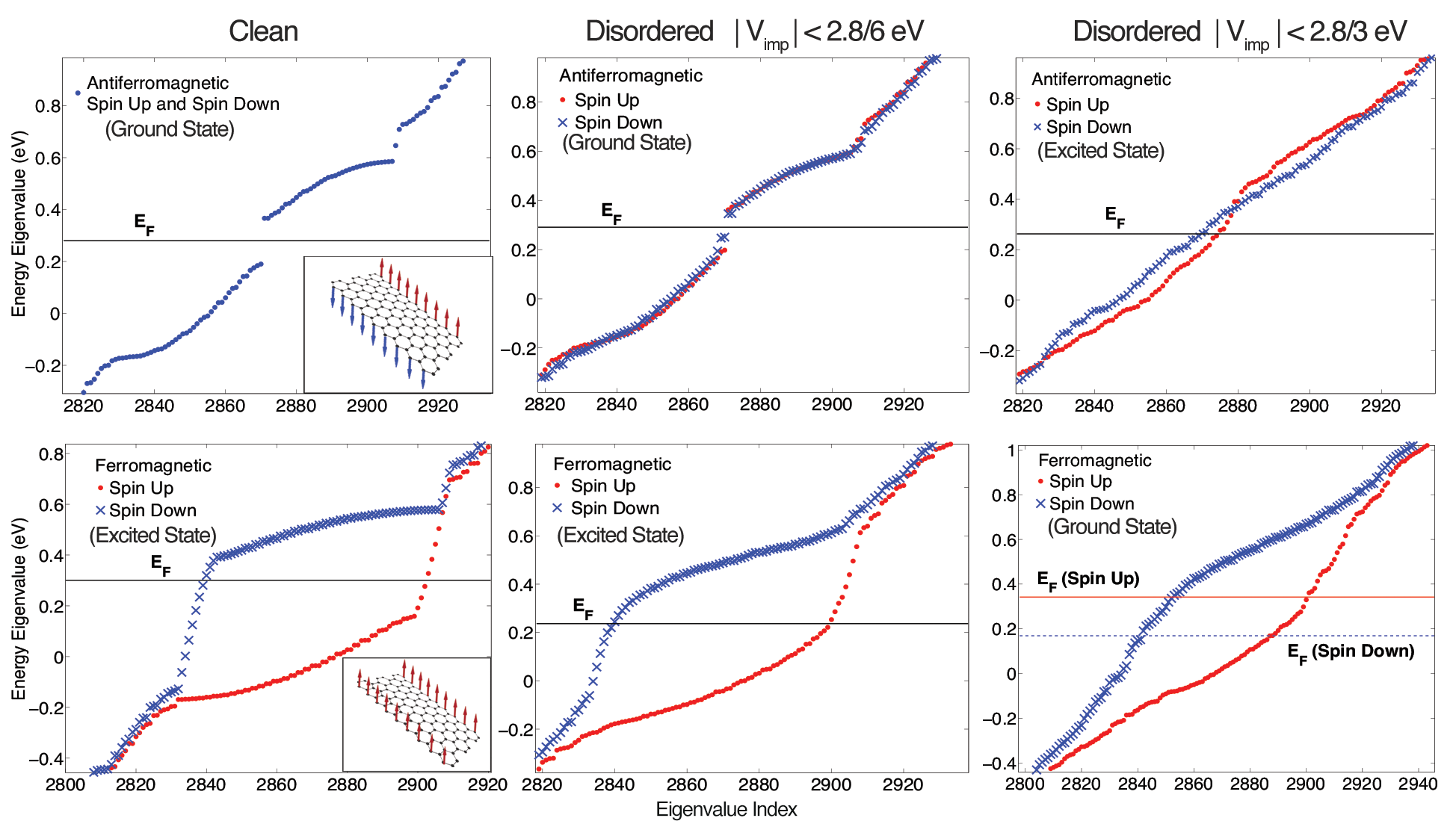}
\caption{(Color online) Mean-field Hubbard spectra for
  antiferromagnetic (top panels) and ferromagnetic (bottom panels)
  phases, for various degrees of disorder strengths, characterized by
  $V_{imp}$. $E_F$ spin up and spin down show the spin-dependent 
  Fermi levels.}
\end{figure*}
Before discussing magnetic properties of the nanoribbons, we first
focus on the combined effect of long-range potential fluctuations and
electron-electron interactions on the electronic properties of edge
states. Fig. 2 shows the electronic density profile corresponding to
the 30 highest occupied valence states (top panels), and the 60
highest occupied valence states (bottom panels), obtained using
tight-binding (left panels) and mean-field Hubbard calculations (right
panels), for the disorder configuration given in Fig. 1. We note that
in the absence of disorder, valence states include about 30 edge
states. In the absence of electron-electron interactions, the main
effect of including disorder is to disrupt the edge states, creating
highly localized edge states. Note that the (hole) edge states
observed in the tight-binding results are not localized in the
p-regions indicated in Fig.1c. Within the extended Hubbard model,
however, electrostatically more correct spin-dependent filling order
of the edge states is obtained, and the hole edge states close to the
Fermi level are now located mostly at the p-regions.  Another
important effect of electron-electron interactions is that the edges
states are recovered within the 60 highest valence states. Thus,
electron-electron interactions makes the edge states more robust
against disorder by partially restoring the symmetry of the system.
Appearance of bulk impurity states is also visible in Fig. 2.  An
interesting question that arises is how the magnetic properties are
affected by the combined effect of disorder and electronic
interactions, which we will be the focus of the rest of this work.

In Figure 3, we show the mean-field energy spectra for
antiferromagnetic (AFM, top panels) and ferromagnetic (FM, bottom
panels) phases, for various degrees of disorder strengths.  When no
disorder is present, the ground state is AFM and the energy spectrum
reveals a gap of the order of 0.17 eV, in agreement with previous
theoretical work \cite{Magda2014,Jung2009,Guclu2013,Fujita} and recent
experimental results\cite{Magda2014}. When disorder is included such
that $\vert V_{imp} \vert < \vert t_{nn}\vert / 6$, the AFM gap is
reduced to $0.1$ eV, and the ground state is still AFM. However, when
the disorder strength is doubled, the AFM gap is practically closed
and the system becomes FM. We note that these results are consistent
with the experimental results of Ref.\onlinecite{Magda2014}, where a
closing of the gap was observed for ribbon with widths larger than 7
nm, which was attributed to temperature and doping effects. Here we
show that, although our system is globally charge neutral, local
formation of electron-hole puddles due to long-ranged potential
fluctuations can also induce a AFM-FM transition.

\begin{figure}
\includegraphics[width=\linewidth]{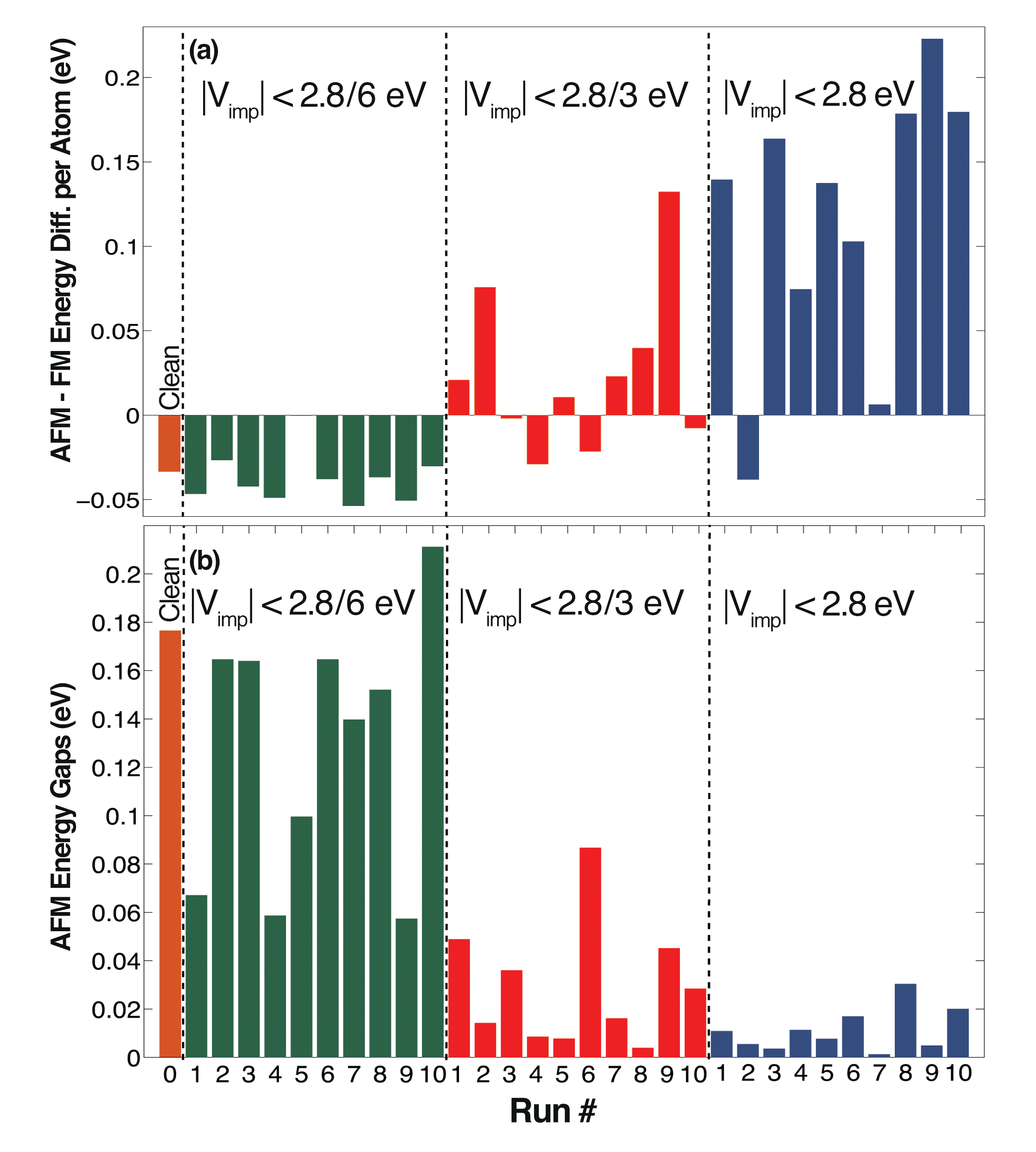}
\caption{(Color online) 
(a) Energy
difference per atom between the AFM and FM phases and (b)
antiferromagnetic phase energy gap for 30 different disorder
configurations with various degrees of disorder strengths.
Strong disorder effect causes system to become
ferromagnetic. For lower potentials, chance of phase transition
reduces.}
\end{figure}

The results of Fig. 2 were obtained for the particular disorder
configuration shown in Fig.1. In order to check the consistency of the
results, we have repeated the calculations for a total of 30 different
impurity configurations and strengths. Figure 4a shows the energy
difference per atom between the AFM and FM phases, a negative value
indicating that the ground state is AFM. For impurity strengths $\vert
V_{imp} \vert < \vert t_{nn}\vert / 6$ no significant effect of
disorder is observed. However, for $\vert V_{imp} \vert < \vert
t_{nn}\vert / 3$, FM phase becomes more dominant. Finally for $\vert
V_{imp} \vert < \vert t_{nn}\vert $, all but one out of ten random
impurity configurations give FM ground state. In Fig. 4b we plot AFM
spectra energy gaps corresponding to the same configurations in
Fig. 4a, showing that the gap quickly decreases as the disorder
strength increases.

\begin{figure}
\includegraphics[width=\linewidth]{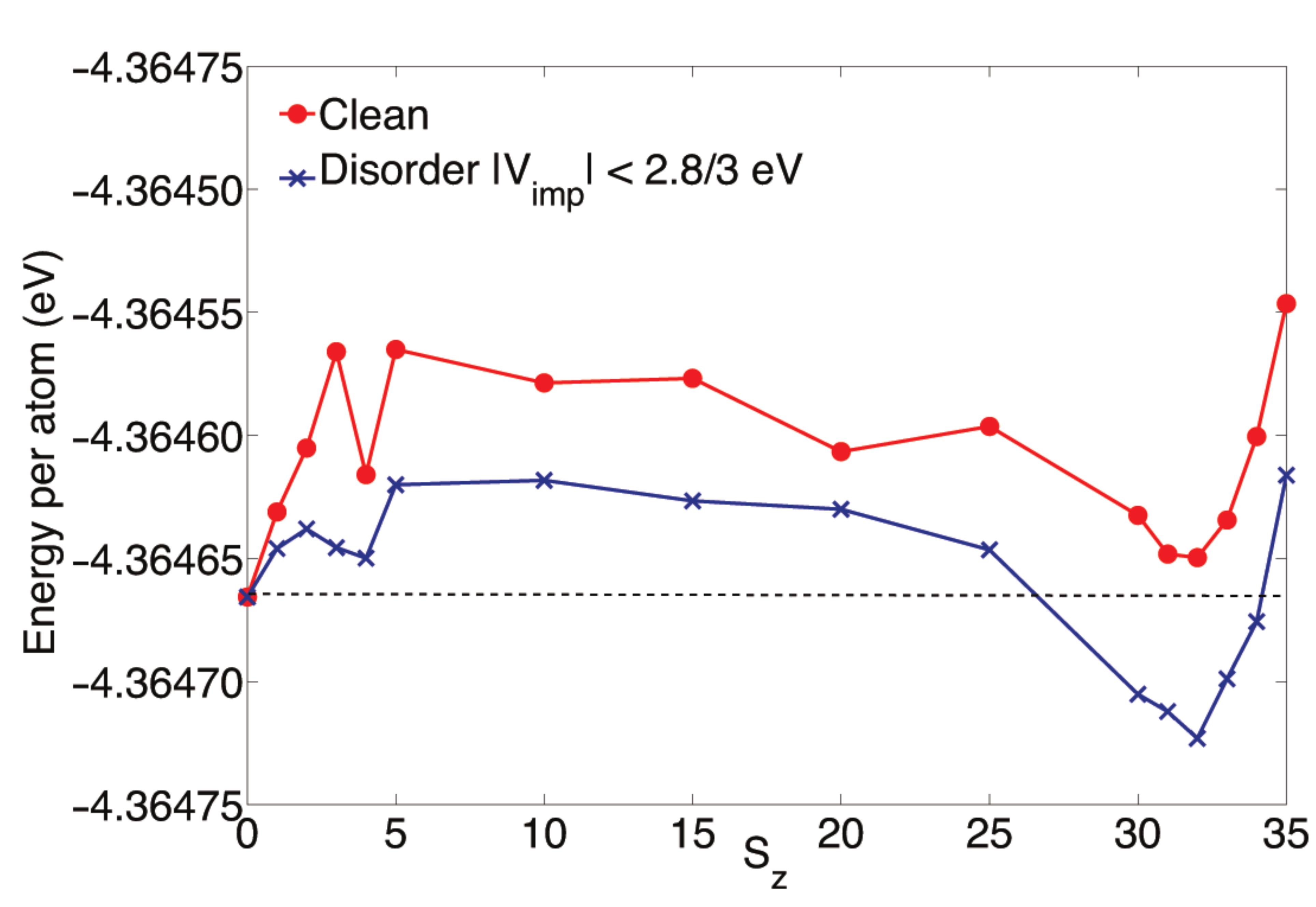}
\caption{Total energy of nanoribbon as a function of magnetization
$S_z$. For clean case, the ground state has $S_z=0$, and for disordered
case $S_z=32$, indicating a FM-AFM phase transition without involving
other possible magnetic phases}
\end{figure}

As discussed earlier, the AFM phase corresponds to $S_z=0$ and the FM
phase corresponds to $S_z=32$.  In order to make sure that no other
magnetic phases (which could be due to the presence of electron-hole
puddles) were not missed in our calculations we have also performed
mean-field calculations for other values of $S_z$ between 0 and
35. Figure 5 shows the total energy of the clean and disordered
nanoribbon as a function of $S_z$, for the disorder configuration
shown in Fig.1b. Clearly, within the mean-field approximation,
the most important magnetic states that dominate the low energy
physics are the AFM and FM phases. We observed similar behavior
for other disorder configurations as well.

\begin{figure}
\includegraphics[width=\linewidth]{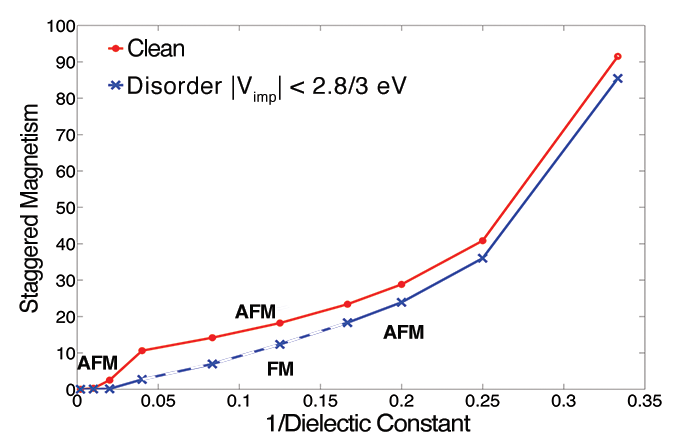}
\caption{Staggered magnetism as a function of dielectric constant
  $\kappa$. Clean system (upper line) shows AFM (solid line) coupled
  edges for all values within 1/$\kappa=[0.33,0.002]$ range. However,
  FM (dashed line) phase transition occurs between
  1/$\kappa=[0.167,0.04]$ after introducing the impurity landscape
  (lower line). For lower $\kappa$ values electronic interaction
  effects become dominant over the impurities hence the system shows
  AFM phase again.}
\end{figure}

Up to this point we performed all calculations with $\kappa = 6$ whose
value determines the magnitude of electron-electron interaction.  As
there are three main energy variables in our Hamiltonian, hopping
parameter, impurity strength and interaction strength, it is also
worth investigating the effect of changing $\kappa$. To see the
interplay between $\kappa$ and magnetism, same calculations are
performed within $1/\kappa=[0.3,0.002]$ interval. A convenient way of
investigating the AFM phase is to use staggered magnetism which is
defined as $(-1)^x(n_{i\uparrow}-n_{i\downarrow})/2$ where x is even
for A and odd for B sublattice sites. In Fig. 6, the change of
staggered magnetism as a function of dielectric constant is shown. For
clean system, no phase transition is observed in this range. On the
other hand disordered system shows FM behavior between
$1/\kappa=[0.167,0.04]$. Recovered AFM phase for $1/\kappa > 0.167$ is
due to strong electron-electron interactions that suppress the effect
of impurities. These results are consistent with our previous results.
For $1/\kappa < 0.04$ region magnetic properties can be neglected.

\section{Conclusions}
To conclude, we have investigated the combined effects of
electron-electron interactions and random potential fluctuations
on the stability of edge states and magnetic phases. The electronic
stability of edge states is found to be surprisingly robust against
disorder due to electron-electron interactions. Moreover, as the
disorder potential strength is increased, the system goes through
an antiferromagnetic to ferromagnetic phase transition, in agreement
with the experimental results of Ref.\onlinecite{Magda2014}. Although the 
possibility of such a transition is well known from previous
calculations\cite{Jung2009} for charged system, here the nanoribbon is
charge neutral. Thus, the magnetic transition is due to the
formation of electron-hole puddles, i.e. local breaking of charge
neutrality.

\section{Acknowledgment} This research was supported by the Scientific 
and Technological Research Council of Turkey T\"UB\.{I}TAK under the
1001 grant project number 114F331 and by a BAGEP grant from Bilim
Akademisi - The Science Academy, Turkey.

%==========================================================================
%                      REFERENCES
%==========================================================================

\end{document}